# "MR Fingerprinting for imaging brain hemodynamics and oxygenation"

T Coudert[1], A Delphin[2], A Barrier[1], EL Barbier[1], B Lemasson[1], JM Warnking[1], T Christen[1]

[1]Université Grenoble Alpes, INSERM U1216, Grenoble Institut Neurosciences, GIN, Grenoble, France
[2]Université Grenoble Alpes, INSERM US17, CNRS UAR3552, CHU Grenoble Alpes, IRMaGe, Grenoble, France

# <span style="color:red">Abstract (101/300)</span>

Over the past decade, several studies have explored the potential of Magnetic Resonance Fingerprinting (MRF) for the quantification of brain hemodynamics, oxygenation and perfusion. Recent advances in simulation models and reconstruction frameworks have also significantly enhanced the accuracy of vascular parameter estimation. This review provides an overview of key vascular MRF studies, emphasizing advancements in geometrical models for vascular simulations, novel sequences, and state-of-the-art reconstruction techniques incorporating machine learning and deep learning algorithms. Both pre-clinical and clinical applications are discussed. Based on these findings, we outline future directions and development areas that need to be addressed to facilitate their clinical translation.

# <span style="color:red">Glossary</span>

**ASL**: Arterial Spin Labeling
**ATT**: Arterial Transit Time
**BOLD**: Blood-Oxygen Level Dependent
**bSSFP**: balanced Steady-State Free Precession
**CA**: Contrast Agent
**CBF**: Cerebral Blood Flow
**CBV**: Cerebral Blood Volume Fraction
**CEST**: Chemical Exchange Saturation Transfer
**DB-SL**: Dictionary-Based Statistical Learning
**DBM**: Dictionary-Based Matching
**DSC**: Dynamic Susceptibility Contrast
**FA**: Flip Angle
**GESFIDSE**: gradient-echo sampling of free induction decay and spin-echo
**GLLiM**: Gaussian locally linear mapping
**HD-MED**: High Dimensional Mixtures of Elliptical Distributions
**HEPI**: Hybrid Echo Planar Imaging
**LSTM**: Long Short Term Memory
**MRF**: Magnetic Resonance Fingerprinting
**MRvF**: MR vascular Fingerprinting
**MT**: Magnetization Transfer
**NN**: Neural Network

**pCASL**: pseudo-continuous ASL
**PLD**: Post-Label Delay
**R**: mean vessel radius
**SNR**: Signal-to-Noise Ratio
**SO2**: brain oxygenation saturation
**TE**: Echo Time
**TR**: Repetition Time
**TTP**: Time to Peak

# Introduction

### The MRF framework

The Magnetic Resonance Fingerprinting (MRF[1]) framework was proposed in 2013 for the simultaneous quantification of multiple parameters from complex MR sequences. At its core, the MRF process involves 3 elements (Figure 1): (1) An MR sequence with time-varying parameters (flip angle (FA), TR, TE, *etc.*) that produces multiple images of the same slice. Although there are no theoretical limitations to the design of an MRF sequence, the resulting temporal signal evolutions (or "fingerprints") in every voxel must vary with the corresponding tissue properties (*e.g.*, T1, T2, off-resonance frequency, *etc.*) that are to be estimated. In order to comply with realistic acquisition times, fast k-space acquisition schemes such as variable density spirals are usually used and produce low-quality raw images with large undersampling artifacts[2]. (2) A simulation tool that produces a 'virtual' fingerprint resulting from the application of the chosen MRF sequence on a 'virtual' voxel with known physical or biological properties. Repeating simulations for different property combinations yields a database (or "dictionary") of virtual fingerprints and associated property combinations. This simulation tool must be complex enough to properly account for the effect of the voxel properties on the MR signal, and dictionaries usually contain several hundred thousand virtual fingerprints or more[3,4]. (3) An algorithm that evaluates distances between temporal signals. For every acquired fingerprint, the property combinations corresponding to the closest simulated fingerprinting (or "match") in the dictionary can thus be assigned to the associated voxel, simultaneously producing multiple quantitative maps.

### Advantages of MRF for relaxometry measurements

The concept paper demonstrated that accurate quantitative maps may be derived from very short acquisitions. This originates from several facts: (1) MRF allows the use of MR sequences in their transient states that produce signal evolutions that are otherwise uninterpretable with analytical models; (2) MRF allows for the simultaneous measurement of physical effects (B0, B1 inhomogeneities, *etc.*) that are usually considered as artifacts in standard quantitative MRI protocol; (3) The dictionary-based analysis (1 set of fingerprints computed once for all) allows the use of complex data models that otherwise take too long to use in a standard fitting task (the model is fitted to the signal from each pixel from each image of each acquisition); (4) The analysis process – all properties are derived from the same train of acquisitions – strongly limits noise propagation effects when compared to the combination of parametric maps originating from multiple sequences, each with its own noise level (*e.g.*, a sequence to map B1 and a sequence to map T2).

Owing to optimization studies on acquisition patterns and image reconstruction, combined with the increasing use of neural networks (NN) to accelerate and improve the dictionary matching process, it is now possible to create parametric maps of T1, T2, and proton density, as well as synthesizing conventional contrast volumes such as T1w or FLAIR images of the whole human brain (1mm$^3$ spatial resolution) with a 1-minute acquisition and 5 minutes reconstruction MRF protocol[5] or other fast protocols[6,7]. MRF protocols have now reached the clinic and can be applied to several organs (*e.g.*, brain, heart, muscles, prostate) and pathologies including stroke, cancer, or neurodegenerative diseases[8].

### Other MRF applications

Despite the initial focus on relaxometry, the MRF framework is not limited to a single excitation pattern (1 single MRF sequence) nor the measurement of the transverse or longitudinal relaxation times (T1, T2) only. It is theoretically possible to quantify any tissue property, as long as the MRF sequence is sensitive to the parameters of interest, provides different fingerprints for different tissue properties, and that the MRF simulations are realistic enough to capture the signal variations. Several teams have extended MRF to the measurements of a large variety of properties. Studies proposed to quantify B0 inhomogeneities and corresponding T2* information using TE variations in the MRF pattern[9,10]. The same parameters have also been assessed with radically different sequences in two recent studies[5], illustrating the flexibility of the approach. Water diffusion coefficients[13,14], and directions[15], musculoskeletal properties[16,17], MR spectroscopy[9,18] or Chemical exchange saturation transfer (CEST) and Magnetization transfer (MT)[19–21] have also been assessed with MRF.

### Limitations of standard MR brain microvascular measurements

Following this line of work, it seemed clear that information about the brain vascular network could also be measured with MRF. It is indeed well known that microvascular architecture and functions can impact MR signals. This is the basis of the Blood-Oxygen Level Dependent (BOLD) effect and many MRI methods using endogenous or exogenous tracers (steady-state or dynamic susceptibility contrast[22], dynamic contrast enhanced[23], arterial spin labeling[24], and quantitative BOLD imaging[25,26] have been proposed to provide measurements of microvessel blood volume fraction (cerebral blood volume or CBV), average vessel diameter and density, permeability, blood flow, vasoreactivity, or blood oxygenation. Yet, a number of technical issues of standard MR perfusion approaches are still under investigation, such as: (1) good results obtained in normal tissues are difficult to reproduce in pathological environments; and (2) estimates tend to be less consistent as spatial resolution increases. Some of these difficulties arise from limitations in the biophysical model, when MR signals are compared to a mathematical model derived from many theoretical simplifications. These approximations are necessary to obtain a model simple enough to suit a fitting procedure. However, they tend to introduce systematic errors in the measurements. In addition, these models are generally derived from observations in healthy tissue and are not guaranteed to be valid in pathological cases.

### Towards MRF microvascular estimates

In the case of vascular measurements, the MRF framework could have major advantages over existing approaches. First, the numerical simulations could consider the contribution of several physiological parameters (vessel diameter, density, volume, oxygenation, flow, *etc.*) simultaneously. Nuisance parameters (*e.g.*, water diffusion, B1 inhomogeneities) could either be measured and incorporated in the numerical model or directly estimated using the fingerprints. This greatly reduces the constraints on acquisition sequences, which so far were often designed to eliminate sensitivity to all but a single or very few parameters of interest. Second, the numerical simulations could provide a reasonable description of the MR signal modulations over a large range of physiologically relevant input values. In many analytical approaches, experimental conditions have to be subdivided in separate regimes (static dephasing, diffusion narrowing, *etc.*) to ensure accurate estimates. These regimes are however not fulfilled in several experimental conditions (*e.g.*, during the passage of a bolus). Third, vascular fingerprinting could provide a real advantage in tissues with irregular vascular networks and pathophysiological abnormalities. Numerical models could be adapted to these irregular vascular networks and accurate parameter estimates would be obtained as long as the corresponding virtual voxels are incorporated in the database. Fourth, new MR sequences with improved sensitivity could be found and used to remove the need for contrast agent injection, potentially improving perfusion map spatial resolution and decreasing the risk of nephrogenic systemic fibrosis and gadolinium retention in brain or kidney tissue[27]. Finally, parameter accuracy would benefit from reduced noise propagation, compared to existing multiparametric methods where multiple maps need to be combined a posteriori.

In this paper, we review the studies that have been proposed in the last ten years to use the MRF framework to measure vascular properties with or without the use of exogenous contrast agents. Specifically, we focused on MRF techniques inspired by steady-state perfusion or quantitative BOLD approaches and designed to measured microvascular network geometrical properties (CBV, R, vessel orientation, *etc.*) or oxygenation saturation (SO2), and MRF methods issued from ASL type acquisition that quantify dynamic blood properties such as cerebral blood flow (CBF) and arterial arrival times. A summary of the studies presenting the major methodological developments is provided in Table 1.

# First MRvF studies in humans

The first study using the concept of MRF for vascular measurements was proposed in 2014[28] and named MR vascular Fingerprinting (MRvF). In this first implementation, the MRF sequence was the "gradient echo sampling of the free-induction decay and spin echo", or GESFIDSE sequence[29], already used in previous quantitative BOLD experiments[25]. The parameter variations were thus limited to echo time changes to produce temporal signal variations acquired with fully sampled Cartesian schemes. Acquisitions were also performed before and after the injection of a contrast agent (ultra-small superparamagnetic iron oxide), and the ratio of the two signals obtained was used as a fingerprint. Contrary to standard MRF used for relaxometry measurements, the simulated voxels in MRvF did not merely consist of a list of properties such as global T1 or T2 values but a sub-voxel structure was generated to represent the microvascular network. Such a structure can be generated by randomly placing straight vessels of a certain radius R until a desired blood volume fraction (CBV) is reached. In this first implementation of MRvF, a 2D representation was chosen, with straight vessels perpendicular to the plane. The blood inside the vessels was assigned an oxygenation (or SO2) value, giving rise to a difference in susceptibility between the blood and extravascular compartments. Synthetic signal evolutions were obtained based on simulations including magnetic field distributions, magnetization evolution, and water diffusion effects[30]. The matching was then determined using least square minimization. Five healthy human volunteers were imaged at 3T and a dictionary of 52,920 signals was eventually generated with three varying parameters, CBV, R, and SO2. The results were also compared to those obtained using more conventional MR methods, steady-state contrast imaging for CBV and R and a global measure of oxygenation obtained from the superior sagittal sinus for SO2.

The general findings of the study were that the MRvF method produced high-resolution parametric maps of the microvascular network with expected contrast and fine detail (even if no information or initialized values were given *a priori*). Numerical values in gray matter (CBV=3.1±0.7%, R=12.6±2.4µm) were consistent with literature reports and correlated with conventional MR approaches. In the gray matter, the SO2 values (SO2=59.5±4.7%) were also in agreement with both the literature from other imaging methods and from the MR-based phase measurement of SO2 in the sagittal sinus. The values obtained in the white matter (53.0±4.0%) were however lower than expected, and the interface between gray and white matter was systematically estimated as highly deoxygenated. Figure 2 shows the MRvF protocol and results obtained in one subject with this approach. Through numerical simulations, the authors[28] identified a probable lack of signal diversity in the generated dictionary. The simplicity of the 2D voxel representation was also pointed out as a limitation and it was already suggested that numerous improvements should be made before being able to apply MRvF to pathological cases. In particular, additional parameters could be considered in the simulations to account for white matter magnetic microenvironment and the realism of blood vessels geometry could be improved. It was also clear that the sequence used for this proof-of-principle study (70 pulses, mainly TE variations) could be changed to a more complex pattern with additional data points and undersampling schemes that could lead to better vascular sensitivity and possibly remove the need for contrast agent injection.

A small follow up study examined if small variation of blood oxygenation could actually be detected with MRvF[31]. Ten volunteers were scanned while breathing different gas mixtures (Hyperoxia (100%O2), Normoxia (21%O2), Hypoxia (14%O2)) for 6 minutes. The MRF protocol was identical to the previous study[28] and a blood T2 mapping sequence based on the TRUST approach[32] was used for independent venous blood oxygenation measurements (SvO2). The results obtained with both physiological monitoring (arterial saturation, SaO2) and TRUST measurements indicated global changes of blood oxygenation during the gas challenges. While the MRvF maps and values obtained during Normoxia were consistent with previous reports, no changes were observed in SO2 estimates between the gas challenges. In order to improve the results, a theoretical study on the dictionary was conducted (Figure 3) and led to the design of a new MRvF pattern based on the same acquisitions but with concatenated pre-and post-signals instead of their ratio. These fingerprints produced similar CBV and R results but SO2 estimates were now significantly different between hypoxic and hyperoxic measurements in the gray matter (Figure 4) suggesting that simple sequence optimization could already improve blood oxygenation sensitivity. The results obtained in the white matter also confirmed that the model used for vascular simulations was probably not complete.

# MRvF in animal brain lesions

As introduced above, MRF for vascular measurements should have a significant impact in pathological environments. In an animal study using multiple models of brain lesions (n=115), three different rat models of brain tumors (9L, C6, and F98) known to present irregular microvascular architectures were analyzed[33]. A stroke model in which the geometry of blood vessels is less affected but blood flow and blood oxygenation present large variations was also included. The results were compared to those obtained from conventional analytical MR methods: steady-state susceptibility contrast imaging for CBV and Vessel Size Imaging[34,35] (VSI) and multiparametric quantitative BOLD (mqBOLD) imaging for SO2 estimates[36].

The authors[33] showed that the fingerprints were sensitive enough to allow the delineation of pathological regions and to distinguish tumor models from one another. In particular, they showed that C6 and F98 glioma models have similar signatures while 9L presents a distinct evolution (Figure 5). These observations could be linked to previous ones, made with histological measurements and other imaging modalities[34–36], showing that C6 and F98 glioma models do not generate new blood vessels for nutrient supply, while it is the case in the 9L glioma model. Good agreements were found between MRvF and conventional MR approaches in healthy tissues and in the C6, F98, and permanent stroke models. Discrepancies were however observed in blood vessel radius estimates in the 9L model. While a possible explanation could be related to the difference that exists between the averaged vessel radius as measured with MRvF and the VSI (defined as a weighted mean), it is also possible that the 9L model contains vascular networks that were outside of the mathematical model limits. In the study, the effects of increasing the complexity of the numerical models were also investigated by (1) adding a new dimension to the dictionary (water diffusion coefficient); (2) increasing the size of the dictionary by extending the range of parameters accessible; and (3) designing special cases where the voxels contain abnormal large blood vessels with preferential orientations. By increasing the size of the dictionary from 40,000 fingerprints to more than a million, the results of MRvF were improved and supplementary information was obtained only by changing the numerical representation of the vascular network. A substantial difference between SO2 estimates remained in the 9L glioma model, further suggesting fundamental difference between both approaches. It thus appeared necessary to further validate MRvF blood oxygenation estimates with other independent measurements.

# MRvF with realistic vascular networks

It is quite clear that the 2D cylinder shape used in early MRvF studies was far from reality. The model was complex enough to compute magnetic field perturbations due to the presence of paramagnetic sources in the blood compartment, but only considered vessels which were either parallel or orthogonal to the main magnetic field. This fails to encompass the variety of orientations and tortuosity of an *in vivo* vascular network (Figure 6), and the resulting interactions of the associated perturbations. In addition, some geometrical properties, such as the tortuosity, are expected to change between healthy and pathological settings.

To obtain more realistic simulations, one simple approach is to perform computations in a 3D volume, in which cylinders with variable orientations are placed. This solution retains the simplicity of generating voxels with defined, controlled CBV and mean vessel radius (R), helping in evenly populating the parameter space of the dictionary. However, the vessels are still mere cylinders without curvature. A second approach is to directly use vascular networks observed in tissues. Using state-of-the-art imaging techniques and data processing, several groups have indeed obtained whole-brain vascular networks from mice, with a microscopic resolution. Some of these datasets are publicly available online[37,38,39] and it has been shown that Fourier transform based calculations can be used to calculate realistic magnetic field deformations[40]. A first attempt to use such realistic structures in the MRvF framework was based on 6 mouse cortex angiograms stored as volumes with 1-µm isotropic resolution[41]. Different geometrical transformations were used for data augmentation to compensate for the small number of animals and for the lack of diversity of brain structures. However, the maps and quantitative results at the group level were not as promising as expected, probably due to a lack of generalization of the dictionary. More recently, 3D voxels segmented from multiple open-access datasets of whole-brain, healthy mice vascular networks, were used as a basis to create an MRvF dictionary[42]. The datasets were chopped to obtain 11,000 MRI-sized voxels and numerical erosion was performed to reach a total of 28,000 different voxels, in which vascular networks were characterized to obtain, for each voxel, a CBV and R value. Note that this gives an uneven coverage of the parameter space.

From these geometrical bases, dictionaries were generated by attributing SO2 and T2 values following a Sobol series as well as a unique water diffusion coefficient of 1,000 µm²/s in each voxel. For the sake of comparison, voxels with the same parameters were generated using 2D and 3D straight cylinders. Experiments were conducted at 4.7T, using a concatenation of GESFIDSE sequences (see Figure 2) and parameter map reconstruction was performed with classical matching as well as a dictionary-based learning approach[43] (see section below on advanced MRF reconstruction). MRI data were acquired in both a healthy and a cerebral 9L tumor bearing group of rats. Optic fiber partial pressure of oxygen (pO2) measurements were performed *in vivo* in the tumoral tissue on the pathological group for SO2 estimates validation.

The authors[44] showed clear differences in the estimated parametric maps between the three different dictionaries used (2D and 3D cylinder-based, and 3D microscopy-based). The SO2 maps indicated the tumor as hypoxic using the 2D dictionary, and hyperoxic using the 3D microscopy-based one, yet the pO2 measurements showed an increased oxygenation in the tumor, as illustrated in Figure 7. The 3D cylinder-based dictionary did not yield a significant difference in SO2 between the tumor and the normal-appearing tissue. This study clearly underlined the importance of using a realistic enough model, as the three dictionaries produced plausible CBV and R maps with similar trends between the tumor and contralateral regions, but gave three clearly different results for the SO2 maps.

Having a more realistic model does not only bring better estimations, but also provides new metrics. A graph-based analysis on voxels containing realistic vascular networks has been proposed to assess characteristics of individual vessels within a volume, which in turn allowed to assign new metrics to each voxel, such as mean vessel length, tortuosity and anisotropy[45]. These features are known to vary between healthy tissue and lesions, especially in tumors. Combining 45,000 realistic voxels, obtained with the help of data augmentation, with varying SO2 values led to a 135,000-entry dictionary, for which Figure 8 gives an example of maps that have been derived. These works on the geometric realism of the simulations are promising, showing improvements in the accuracy of the measurements proposed in earlier MRvF studies and opening new possibilities. It is worth noting that due to the dictionary matching principle used in MRvF, obtaining these new metrics is only a matter of dictionary preparation (better vessel network characterization). As such, it does not necessitate new simulations nor longer reconstruction times.

## MRvF without contrast agent injection

Although previous MRvF studies relied on acquisitions with the injection of contrast agents (CA), these agents are contraindicated for certain patient populations, or prohibited in some areas of the world such as the Fe-based P904, limiting their use in clinical practice. Recent works have used MRvF to image brain microvascular networks without the need for contrast agents to bridge the gap with clinical application.

Achieving accurate and reliable parameter quantification without contrast agents is challenging due to the reliance on intrinsic tissue properties and to the complexity of accurately modelling and extracting vascular parameters. One MRF approach is to use a spoiled gradient-echo sequence in conjunction with a multi-compartment model (T1+T2) to estimate CBV combined with intravascular water resonance time to assess blood-brain barrier integrity without the need for CA[46]. However, relying on a multi-compartment model also raises the question of the blood compartment homogeneity, as it is known that blood relaxation highly depends on blood hematocrit and oxygenation[47,48]. Thus, a unique couple of relaxation times is not sufficient to capture the whole blood network, and a possible intersection with another tissue component can be foreseen. Relaxation times are also influenced by other factors including water diffusion or blood flow, making it even more difficult to isolate the vascular component's contribution to the signal resulting from a spoiled gradient-echo sequence.

More recently, a new sequence design for MRvF experiments relying on a balanced gradient-echo sequence has been presented (Figure 10 left panel)[49]. By exploiting the inherent sensitivity of balanced steady-state free precession (bSSFP) sequences to the BOLD effect, and leveraging a new simplified description of intra-voxel frequency inhomogeneities, the MRvF-bSSFP sequence aims to simultaneously measure relaxometry (T1, T2, and PD), magnetic fields (B1, B0) and microvascular parameters (CBV, R) in the human brain without contrast agent injection. Here, the sequence was designed with limited SO2 sensitivity for which the value is fixed at 60% in the model. The first results on a cohort of 6 healthy volunteers agreed with reference MRF sequences for standard relaxometry measurements, whereas microvascular estimates need to be properly validated against contrast-agent acquisition but are in line with previous studies, especially for the gray/white matter CBV ratio which was estimated at 1.91. The main limitation of this work is the computational requirements for estimating a large number of parameters and therefore for generating billion-entries MRF dictionaries. For sequences with hundreds of time points, it means a storage capacity of more than a terabyte to store the dictionary and an incredibly time-consuming matching process implying memory-greedy operations. To tackle this, batched dictionary matching was performed using sub-dictionaries, simulated on-the-fly during the reconstruction process and accounting for only a few intra-voxel distributions. In the end, the best-matched result among all sub-dictionaries was kept for each *in vivo* fingerprint. A schematic illustration of this process is shown in Figure 9. This innovative MRF reconstruction method reduced memory requirements but still required long reconstruction time, which could limit clinical applications.

# Dynamic Susceptibility Contrast MRF

While the first MRvF approaches have been based on steady-state perfusion and quantitative BOLD approaches (*i.e.*, static characteristics of microvascular networks), a recent study[50] has applied the same principles to analyze MR signals during a Dynamic Susceptibility Contrast (DSC) experiment and quantitatively characterize microvasculature in 6 patients with gliomas. The hybrid echo planar imaging (HEPI) sequence, a recent technique to merge the acquisition of GRE & Spin Echo (SE), was played during the first passage of a Gadolinium-based contrast agent.

The exact properties of the HEPI sequence (as played out on the scanner) were imported into a Bloch-based Dynamic Contrast Enhanced (DCE) simulation tool that also takes into account contrast agent extravasation and diffusion[30]. The T2* (GRE) or T2 (SE) signal variations during CA bolus passage were used to construct dictionaries in which vessel permeability (k), mean vessel radius (R), and relative CBV (rCBV) were varied. After standard matching, the vessel parameters were also compared to those quantified using a conventional VSI technique. Both techniques yielded similar rCBV maps for the gray and white matter in all patients, and clearly differentiated the lesions. For rCBV, a significant quantitative difference was obtained in a patient with an enhancing glioma. Vessel radius maps obtained from VSI showed lower values compared to those from MRvF, which could be explained by VSI underestimation of larger vessel sizes. It was also noted that the MRF approach was more resilient to noise and CA leakage.

A major constraint of the DSC-based MRvF is that a temporal resolution of 1.5–2 sec is needed, which limits the spatial resolution as well as the coverage that can be achieved. Yet, a major advantage of this approach is that, by scanning during the dynamic bolus passage, the concentration of the contrast agent, and thus the magnitude of magnetic susceptibility effects, covers a wide range and potentially yields more information on microvasculature than traditional MRvF. While the number of vascular parameters used to simulate the signal was limited in this first study, the simulation and dictionary could be extended to include more vascular parameters like blood flow and oxygen saturation, which could improve the matching in the tumor area. Further research could also focus on the extent to which the inclusion of permeability into the dictionary avoids confounding by CA leakage.

# Arterial Spin Labeling MRF

Like dynamic susceptibility contrast, arterial spin labeling (ASL) is a technique sensitive to the dynamics of the delivery of blood through the vascular tree and to the tissues. ASL probes perfusion without administration of exogenous contrast agents by modifying the longitudinal magnetization of water protons in the arteries using RF pulses, thus creating a bolus of labeled water upstream of the organ of interest. In contrast to other perfusion MRI techniques based on the administration of contrast agents such as Gadolinium chelates, which remain intravascular in the healthy brain and have a clearance on the order of hours, the ASL label diffuses freely in tissues and is of short lifetime, decaying with the T1 of blood water over the course of seconds. This offers the possibility of rapidly repeated measurements, giving access to functional imaging of perfusion, and renders ASL mainly sensitive to the arterial part of the vascular tree. Key parameters accessible to basic ASL signal models are CBF and arterial transit time (ATT)[51]. Additional parameters such as the permeability of vessels to water, and intravascular water volume can be assessed via two-compartment models[52].

Many ASL acquisition schemes exist, varying in the way the arterial water is labeled and in the sampling of the signal. Today, the most widely used labeling method is pseudo-continuous labeling[53] (close to the originally proposed continuous labeling[54]), which inverts the magnetization of blood water by flow-driven inversion as the blood crosses a labeling plane intersecting the upstream arteries. Using a labeling period on the order of seconds, labeled blood water accumulates in the perfused tissues downstream, modifying the overall magnetization ("label condition"). After a post-label delay (PLD), again on the order of seconds, allowing labeled blood to travel to the tissue, an image is acquired. The subtraction of an image acquired without labeling ("control condition") removes the contribution from static spins, leaving a signal proportional to the amount of blood that has perfused the tissue during the label duration. Importantly, sequences are designed to precisely match signals from static spins in the label and control conditions using pre-saturation pulses in the imaging region, background suppression using inversion pulses, and label pulses designed to compensate for magnetization-transfer effects. Temporal characteristics of the label can be extracted using measurements at multiple delays after the label period or by varying between label and control conditions in a monitored manner prior to imaging[55].

While ASL has enjoyed increasing success in clinical imaging, especially since a consensus on an optimal protocol for clinical imaging was elaborated by the ISMRM perfusion study group and the European ASL initiative in Dementia[24], the limited signal-to-noise-ratio and long acquisition times required to perform multi-parametric studies, largely owed to the long label and post-label periods, remain an issue. Like for other quantitative MRI techniques, analysis of ASL data has mostly relied on analytic inversion of signal equations, which assume full delivery of labeled blood and mandate long post-label delays. Additional parameters required for the signal modeling, such as tissue T1, are either measured separately or obtained from literature.

Thanks to the flexibility offered by the generation of the arterial label using RF pulses, independent of injections or external intervention, a sequence and data processing design for ASL in the framework of MR fingerprinting has recently been proposed[56,57,58]. ASL fingerprinting can be viewed as an extension to time-encoded ASL, where label and control periods of varying duration alternate. In contrast to time-encoded schemes, the signal is sampled much more frequently, at the end of each label or control block, either without post-label delay[57,58] or after a variable PLD[56] (Figure 11), and timings in the label and control conditions do not match. Instead of subtracting label and control signals, the effects of the sequence on the signals of static and labeled spins are modeled, taking into account the accumulated contributions of several preceding label and control periods. The shorter label periods and absence of post-label delays lead to a drastically increased frequency of signal sampling with respect to the traditional approaches. The initial report used an average TR of 380 ms, which is more than ten times shorter than typical TRs in single-delay pCASL. This offers the potential of a substantial increase in SNR and provides sensitivity to bolus arrival time. The signal from static spins is explicitly modeled with tissue T1 being one of the model parameters. A sinusoidal variation of label duration / TR over time has been shown to provide optimal precision and accuracy for the estimation of most model parameters[59].

Since ASL is a quantitative technique, parameter estimation is based on a quantitative signal model. Non-fingerprinting ASL data are most often processed using a one-compartment model of perfusion, where complete and immediate exchange of water between the intravascular and extravascular compartments is assumed[51]. In ASL fingerprinting, label or control periods immediately precede the signal acquisition, leading to the presence of ASL signal from arterial blood that does not contribute to perfusion of the tissues inside the voxel. The ASL signal model is therefore adapted accordingly to accommodate pass-through arteries, with additional associated parameters of arterial bolus arrival time, arterial volume and blood travel time through the artery in the voxel[58]. Additional parameters like label dispersion and magnetization transfer effects can also be included[60]. The flexibility of the MRF framework further allows the incorporation of the MRF-ASL label scheme into MRF sequences sensitive to other parameters. A sequence sensitive to perfusion, diffusion and T2* relaxation has been proposed[61].

MRF-ASL is still a fairly recent method, and so far most studies have focused on optimization of the method. Despite this fact, applications in stroke[14] and Moyamoya disease[62] have already been presented (Figure 12). and initial comparisons to other perfusion imaging techniques like Look-Locker ASL[58], single-delay pCASL[14,62], multi-delay pCASL[60] and DSC[59,62] have been performed.

A major challenge in ASL fingerprinting is the accurate modeling of the effect of the excitation RF pulses in the acquisition modules. These RF pulses attenuate the magnetization of tissue spins, both static spins and spins arrived previously by perfusion, reducing their contribution in later images. Since the modeling of the bolus arrival time (both in tissue and feed-through arteries) requires sampling signals after varying label/control patterns, the excitation RF pulses need to leave a sufficient portion of the longitudinal magnetization in the imaging region intact. The excitation RF pulses provide sensitivity to B1+ and tissue T1, but also attenuate the magnetization of blood contained in pass-through arteries, reducing the available ASL signal for down-stream voxels. In the initial report, a single-slice acquisition with 70° FA was used to avoid this issue, later extended to 7 slices with 40° FA[59]. Full modeling of the impact of excitation RF pulses on longitudinal magnetization of intra-vascular spins in pass-through arteries that will perfuse downstream brain regions in later TRs is extremely challenging. The approach of using in vivo data in a dictionary may alleviate the issue to some extent, but for now it is difficult to assess to what extent this issue has an impact on the accuracy of MRF-ASL.

## Advanced vascular MRF reconstruction

Designing more efficient fingerprints and increasing the size of dictionaries eventually lead to longer matching procedures. Therefore, many efforts have been made to reduce post-processing time in MRF and some advanced reconstruction techniques have been applied in vascular MRF approaches. Dimension reduction techniques have been proposed to alleviate computational costs of image reconstruction, paving the way for more efficient and scalable clinical applications. Standard compression methods, such as SVD, have been successfully applied to spoiled MRF sequences[63], reducing dictionary size by projecting signals into a low-dimensional subspace without compromising map quality. However, sequences like bSSFP, used in non-contrast MRvF, demand a larger number of bases for accurate signal projection[64] which limit the compression impact. Efficient learning of a large-scale dictionary representation can be achieved using compression based on online High Dimensional Mixtures of Elliptical Distributions (HD-MED) with a divide-and-conquer strategy[65]. Validated on healthy volunteers against standard dictionary matching in the origin space, this technique opens the way to efficiently manage MRvF sequences that deal with a large number of estimated parameters.

The development of machine learning (ML) algorithms for MR data reconstruction has also shown interesting results in MRF studies. For MRvF data, Boux et al.[43] introduced a dictionary-based statistical learning (DB-SL) approach. Specifically, this method employs a Gaussian locally linear mapping (GLLiM) model[66] to learn a mapping from fingerprints to parameters while providing a full posterior distribution for each fingerprint. This enables the computation of both parameter estimates and confidence indices. The DB-SL approach also leverages a quasi-random sampling strategy to efficiently generate an informative dictionary. It was compared against both the standard dictionary-based matching (DBM) method and a dictionary-based deep learning (DB-DL) method using a fully connected network. Results from simulations and real data acquired in tumor-bearing rats demonstrated that DB-SL yielded more accurate estimates than DBM, avoiding limitations tied to dictionary boundaries. Notably, DB-SL proved robust to higher noise levels and offered confidence indices for estimates at no additional computational cost. Moreover, the GLLiM algorithm has also successfully been combined with the HD-MED approach[65], aiming to improve the continuity of the estimates and the speed of reconstruction. An innovative reconstruction method has been proposed for high-dimensional vascular MRF approaches using Long Short-Term Memory (LSTM) networks in a bi-directional way[67]. The network, trained with online batches of MR signals simulated for the MRvF-bSSFP sequence[49], could directly map 6 tissue parameters (T1, T2, B1, B0, CBV, and R) with the acquired signal fingerprint. This allowed a second-level reconstruction of six MRvF maps with low computational demand and denser sampling of the parameter space. In addition, the latest results showed that the model is robust to undersampling artefacts.

In MRF ASL, data analysis in the first report was based on dictionary matching[58]. The authors describe a dictionary requiring 171 GB of disk space for the two-compartment (7-parameter) model[58]. Like for other MRF sequences, ML approaches have thus been used to analyze the fingerprinting data. Using a fully connected neural network did not only speed up the matching procedure by several orders of magnitude, but indeed improved the quality of the parametric maps produced[68]. The fully connected network achieved generally better coefficients of determination than dictionary matching when reconstructing parametric maps from simulated noisy data, as well as better reproducibility observed in data from subjects scanned three times each. The authors[68] attribute this to the fact that a neural network can reconstruct continuous parameter maps, and to the robustness to noise acquired by the network in training on noisy data. The accuracy of the reconstructed CBF values from simulations shows room for improvement, though, with a slope of the linear regression between fitted and ground truth values of only 0.7. It is acknowledged that further validation of the accuracy of the parametric maps is necessary[68].

Dictionaries in MR fingerprinting are not necessarily obtained only based on simulations. Aiming to improve the accuracy of parameter reconstruction, Fan and co-workers[69] trained a neural network on a dictionary derived from high-SNR signal time courses acquired *in vivo* by averaging data from ten repetitions of the MRF-ASL protocol in each of 15 young healthy subjects. The rationale of the approach is that the *in vivo* time courses represent more realistic signal evolutions, taking into account effects that cannot be fully simulated, owing to the complexity of perfusion kinetics. The corresponding parameter values are determined by dictionary matching of the high-fidelity time courses. In a second step, neural networks are trained on the dictionary derived from the in vivo MRI data, one separate network for each parameter. The authors[69] found that reconstructing MRF-ASL data using NN trained on high-fidelity acquired data outperformed the other methods (NN trained on synthetic data, dictionary matching) in terms of reliability, increasing SNR of CBF maps by 173% with respect to conventional dictionary matching. Despite the fact that the NN was trained on data from healthy subjects only, the capacity of MRF-ASL and NN reconstruction to characterize perfusion abnormalities in a Moyamoya patient could be shown.

# Future Directions

The vascular MRF approaches bring the hope for fast, multiparametric, quantitative estimates of blood dynamics, microvessel geometry and functions. Better standardized maps would lead to better patient follow-up and efficient multi center studies that do not require post processing harmonization tools. The MRF framework would also facilitate automatic analyses of the data given that all images are also already co-registered and thus would not require lengthy pre-processing steps.

For the moment, the different vascular MRF approaches have shown encouraging results in proof-of-principle studies in human volunteers and rodents. They are well supported by numerical investigations and are issued from already accepted MR perfusion techniques. After several years of technical developments, interesting results have also been obtained in various types of animal models of pathologies and first results have been obtained in small cohorts of patients (Moyamoya, brain tumor patients) and animal models. These methods are now in validation stages where they are mostly compared to more traditional MR techniques (DSC, single- and multi-delay ASL, steady-state perfusion and oxygenation approaches). The main difficulty is that these techniques cannot always be considered as gold standards as they have their own limitations. This is especially true if the underlying models do not take into account certain physical effects. Estimates of new vascular parameters such as network anisotropy, density or blood oxygen saturation also require new types of ancillary techniques but reference measurements are not easy to obtain.

Thanks to the flexibility of the MRF framework, there are several ways to improve the current vascular MRF techniques. As already shown in this review, this can be done in each of the MRF steps, including MRF sequence design, simulations, image acquisition and reconstruction and dictionary matching. For better acceptance in clinical environments, it seems that optimization of acquisition and reconstruction times should be a priority. For example, numerous studies using optimization algorithms for the design of MRF sequences have shown that results from pseudo-random sequence parameters (as used in the firsts MRF papers) can be largely improved in terms of accuracy and acquisition time by maximizing the discrimination between different tissue types. Recent approaches include pattern optimization algorithms using a fast cost function evaluation[70], BLAKJac for transient-state sequence optimization via finite-difference Jacobian evaluation[71] and automatic differentiation of Bloch equations for efficient joint optimization of flip angles and TRs[72]. Advanced methods include OPTIMUM[73], a mix of automating and brute force sequence selection, or SQP-based approaches with B-splines interpolation[74]. The choice of an optimization algorithm is clearly related to the need of the user depending on the dimensionality of the search space, and moreover, the time to evaluate the parameters. The design of the cost function seems to highly influence the results and simple evaluation models have been developed based on MR signal dictionary evaluation[75,76,72] or low-rank subspace evaluation[77]. These techniques could be particularly useful for vascular MRF approaches, as it has been seen in this review that most of the sequences were close to basic standard MR sequences. This is the case for the MRvF approach with contrast agent injection which only contains two FA values and changes in echo times. For contrast-free MRvF, some manual optimization studies have already been explored to increase sensitivity to a small number of parameters[78,79]. The non-contrast bSSFP approach has more parameter variations but still shows a constant TR, linear FA variation, and a single echo time. Small adjustments of these parameters could already impact the vascular results.

On top of MR pattern optimization, it is possible to improve MRF results by working on the image acquisition and reconstruction processes. In general, MRF k-space sampling schemes have been based on spiral and radial designs. Advanced 3D-optimized patterns with variable density schemes have allowed significant reduction of acquisition times[80] and it is clear that vascular MRF methods will benefit from these types of approaches. The reconstruction techniques for undersampled images have also been largely improved over the years using AI-based strategies with deep learning architectures. These networks, like Auto-Encoders[81], could also be a promising perspective for non-contrast MRvF approaches. Parameter estimation with matching algorithms is inefficient in MRvF due to the high dimensionality of the dictionaries and should be replaced by quasi-immediate DL estimators like LSTM[67]. From convolutional networks to auto-encoders[82,83,84,85,86], several other DL structures have proven to be efficient in MRF with reduced aliasing artifacts. All these multiple reconstruction steps could also be done jointly. Recent deep learning architectures such as deep image priors[87] introduced efficient reconstruction techniques of undersampled

MRF acquisitions which are self-supervised, that is, they don't use training on external datasets. These techniques have even been adapted to 3D MRF[88] with non-prohibitive reconstruction time. Due to the difficulty of acquiring vascular datasets, those unsupervised methods are clearly adapted to vascular MRF approaches. It is also important to note that the ML models described so far are voxel-wise models which map the MRF signal of each voxel independently to its bioparameters without incorporating potential spatial links across voxels. Therefore, they have limited capacity in addressing spatially-correlated aliasing/undersampling artifacts deriving from accelerated acquisitions. Other types of networks have been proposed to specifically address this issue for MRF reconstruction and jointly process image voxels, often using anatomical maps or images during training to account for spatial voxel interdependencies[89–91].

Another potential way to improve vascular MRF methods is to further improve the simulation models. MRF hinges on the fact that simulations accurately reflect the biological and physical processes underlying the signal acquisition. It has been shown that the geometrical vascular structures play a large role in the parameter estimates and that the addition of physical effects such as water diffusion significantly impact the results. A solution balancing the need for realism with low data availability could be to take advantage of powerful synthetic vascular networks generation algorithms to create voxels suiting the needs of the population studied by MRI[92,93,94]. Different sources of magnetic susceptibility should also be considered in the tissues. Indeed, only deoxyhemoglobin was taken into account in past MRvF studies, while myelin and iron deposits are also present in brain tissues[95]. In particular, myelin represents a large fraction of the white matter and MRvF models have led to incorrect blood oxygenation measurements even with advanced 3D geometries[96]. A proper representation of the geometry of the myelin fiber will have to be found, as those fibers are much smaller than the vessels considered so far. It is likely that simulation methods relying only on intra-voxel magnetic inhomogeneity distributions instead of a fully 3D-resolved process[49], will circumvent the prohibitive computation time that comes with such high spatial resolutions. It is also interesting to note that similar observations have been made in quantitative BOLD studies and that combination of magnitude and phase (quantitative susceptibility mapping) analyses have been proposed to resolve these issues[97] suggesting that exploiting a complex signal in vascular MRF would likely provide additional useful information. Finally, the models could also be improved by taking dynamic processes into consideration. Considering realistic blood flows in the micro vessels with different arrival times could not only improve MRF DSC results but also benefit MRF ASL. It has also been suggested that these types of models could provide additional measurements of the maximum achievable oxygen extraction fraction in a voxel[98]. Regarding MRF ASL in particular, it remains to be determined which effects among magnetization transfer, intra-voxel transit time, and possibly others, need to be taken into account to maximize the reliability of perfusion parameter estimates.

After optimization and proper validation steps, the vascular MRF techniques could be used to study brain tumors using high-resolution vascular maps in combination with other relaxometric parameters to try to predict recurrence after treatments. The same protocol could be helpful to study subtle vascular variations in neurodegenerative diseases. This is also true for MRF ASL, which has so far been applied in stroke[61] and steno-occlusive disease[62], neurodegenerative diseases and neuro-oncology being the other major clinical applications that have been found to benefit from ASL MRI[99]. These techniques could also be effective in other parts of the body. For example, non-contrast CBV and R measurements could be used for the study of kidney vascular networks where the method could benefit from larger blood signals and therefore higher SNR during acquisition. Given the sequence design flexibility, an optimized acquisition could also open the door for estimating other parameters of interest such as advanced oxygenation measurements which could then be used to study tissue Hypoxia and redefine stroke penumbra beyond the simplified concept of perfusion/diffusion mismatch. An important point to take into account in clinical settings is the treatment of patient motions which has not yet been addressed in vascular MRF studies. Indeed, several parameters rely on measurements of the intravoxel magnetic field heterogeneity, whose estimate may be altered in case of motion. Although the initial MRF work has demonstrated tolerance to motion with specific motion paradigms, additional algorithms, based on motion estimation extracted from MRF measurements, have been developed for 2D MRF[100] to further improve the motion robustness. Furthermore, studies have used sliding-window reconstructions to reduce MRF images aliasing artifacts, and directly perform motion correction in k-space[101]. Prospective motion correction using 3D fat navigators[102] for 3D MRF, and strategies based on residual learning in fast 3D whole-brain multiparametric MRI[103] could also be considered for vascular studies.

# Conclusion

MR Fingerprinting has proven to be a powerful framework for quantitative MRI. With dedicated numerical simulations and adapted MR sequences, it also allows estimates of brain hemodynamics, vascular geometry and blood oxygenation. Advancements in sequence design, microvascular models, dictionary compression, and deep-learning-based reconstructions have even enabled vascular measurements in humans without the need for contrast agent injection and new types of vascular quantitative assessments are still being considered.

Yet, most of the vascular MRF approaches are still in the development stage and need robust validations. While standardized phantoms, such as the ISMRM/NIST can be used for relaxometry techniques[104], or the QASPER phantom for ASL[105], can be used for relaxometry techniques, it is difficult to find such calibrated objects for microvascular estimates. It might however be interesting to take advantage of the strong link that has already been made between vascular MRF and high-resolution microscopy for microvascular structure modeling and dictionary generation. Indeed, light sheet microscopy, phase contrast X ray imaging or photoacoustic measurements could also be used in animal MRF studies to directly compare matched virtual voxels with actual underlying vascular structures. This could further be used to assess the precision of the measurements as well as a means to optimize the techniques.

Some MRF vascular approaches, such as ASL, are more advanced than others and are going through multi-center studies to confirm the reliability of the measurements across different scanner platforms and patient populations. The challenges here, such as low SNR, motion artifacts, and the need for fast reconstruction tools are also being actively addressed in other vascular MRF approaches. With clinical implementation of these techniques, and subsequent acquisition of large datasets in patients, there could also be an opportunity to rely less on simulation and minimize the upfront assumptions. In contrast to simulation-based methods, data-driven approaches could be used to maximize tissue contrast while minimizing acquisition time. Such an approach has been proposed by Fan et al.61 for MRF-ASL acquisition and could be easily extended to MRvF approach. Another data-centric strategy would be to analyze large databases of MRF signals from both healthy and pathological subjects, with known microvascular alterations, using unsupervised statistical clustering techniques. This type of approach has shown promise in the automatic and unsupervised identification of non-physiological tissues from healthy tissues using a limited number of MRI images. Therefore, it seems highly relevant to apply these methods directly to acquired MRF signals, which inherently contain a much richer reservoir of information.

Regarding key applications, Vascular MRF could be useful in studying cerebrovascular diseases, including stroke, tumors, and neurodegenerative disorders. In the acute stage, it provides speed. Then, one can benefit from the quantitative maps to finely monitor the patient evolutions. Moreover, expanding its application to other organ systems, such as cardiac and renal imaging, could also provide valuable insights into systemic vascular health. A great feature of the MRF approach is its flexibility in terms of sequence designs and measured parameters. Thus, the developments reported in this review for microvascular estimates could also benefit other studies on the microenvironment based on diffusion or CEST types MR acquisitions. It might even be conceivable to combine them to offer a more comprehensive view of tissue physiology and new types of virtual MR histology exams.

# Figures and tables

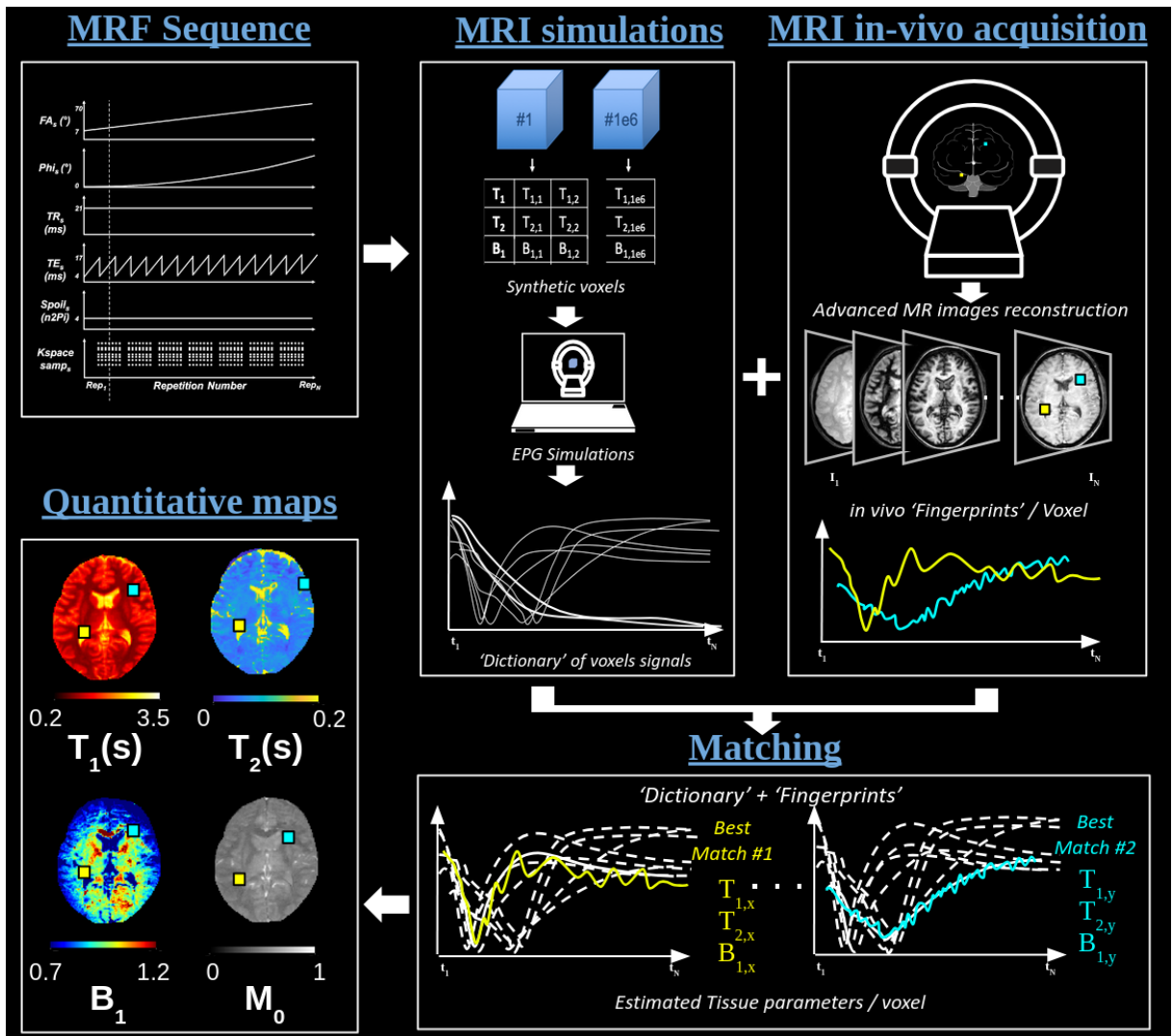

**Figure 1:** MRF framework (FISP sequence for relaxometry measurements) with results on one human volunteer obtained at 3T. The MRF sequence is input in parallel to an EPG-based MR simulator for dictionary generation and to the MRI machine for data acquisition. Acquired "fingerprints" are then matched to the MRF dictionary for each acquired voxel. The voxel-wise reconstruction results in simultaneously computed quantitative maps of T1, T2, B1 and proton density (M0) parameters.

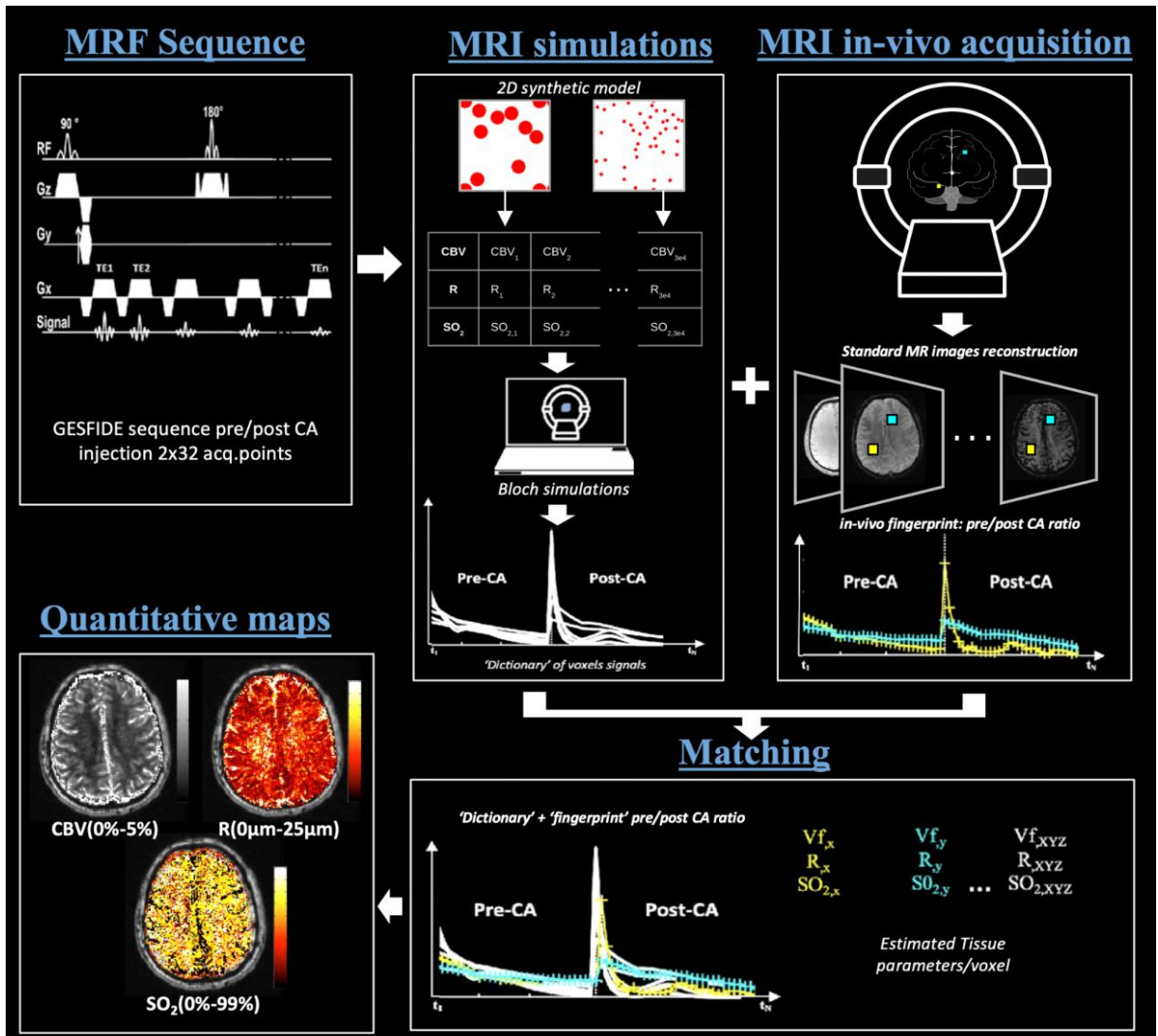

**Figure 2:** Initial MRvF framework with results on one human healthy volunteer obtained at 3T. Note the sequence and simulation tool difference with Figure 1. The MR simulator is Bloch-based and takes into account the effect of the contrast agent, microvascular structures and encoding for CBV,R and SO2 variations. The acquisition protocol includes contrast injection. The matching process allows microvascular properties and quantitative maps of oxygenation to be reconstructed.

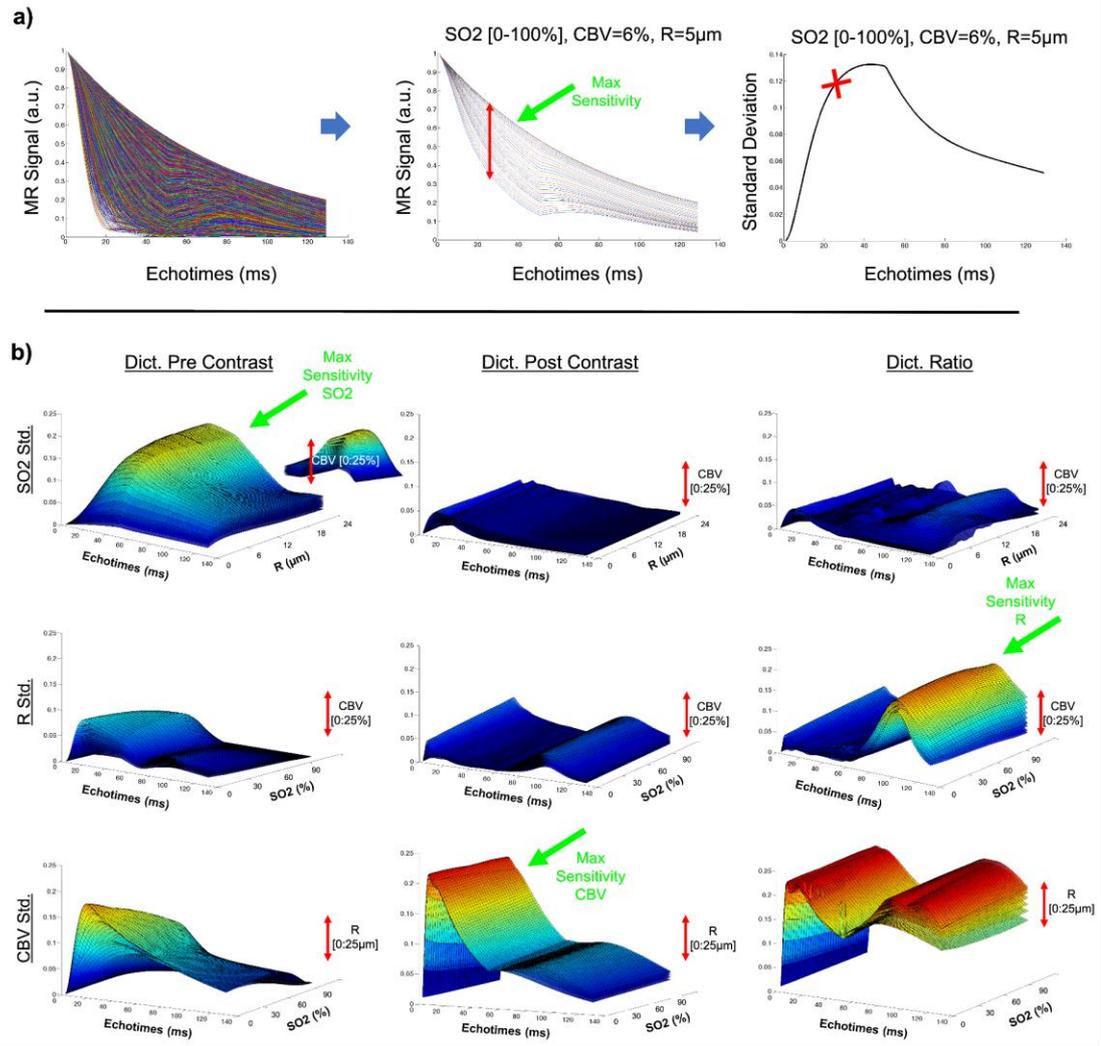

**Figure 3:** MRvF Dictionary study: (a) standard deviation of the signals at each echotime for a particular set of fixed and open parameters indicate theoretical maximum sensitivity. (b) In this example, max sensitivity for SO2, R and CBV don't occur at the same echotimes and depend on the presence of a contrast agent. In particular, SO2 sensitivity is only high in the pre-contrast dictionary. (**adapted from Christen 2018**[31])

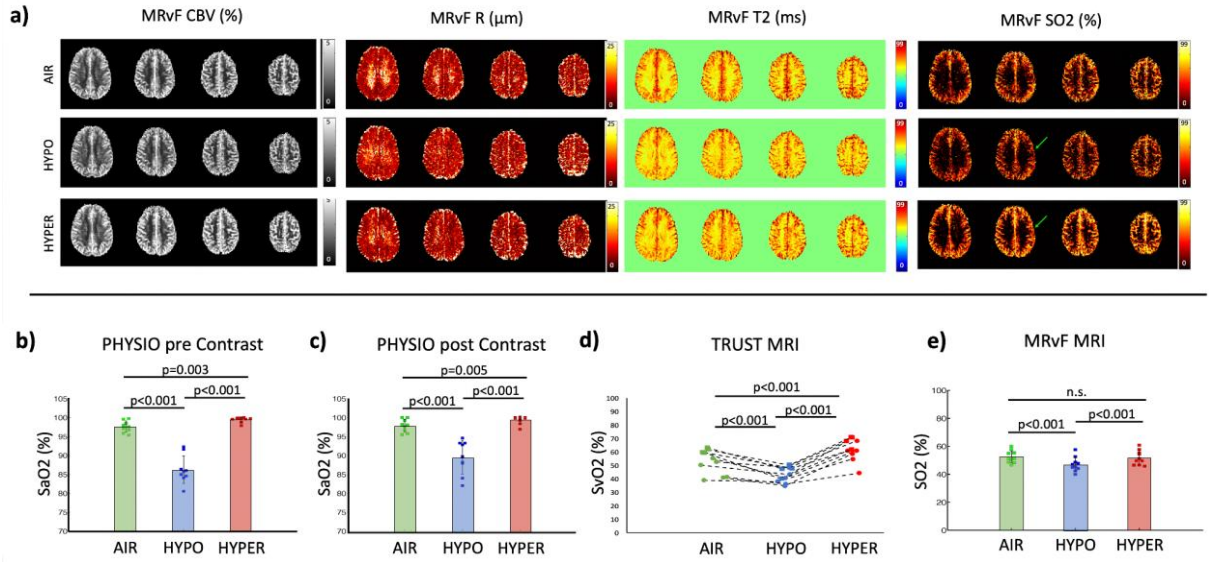

**Figure 4:** MRvF Gas Challenge study in healthy volunteers. (a) parametric maps obtained in one volunteer during AIR, HYPOXIA and HYPEROXIA challenges. Gray matter changes can be observed in the blood oxygenation SO2 maps between hypo and hyper conditions. Bottom: Group results (n=10) obtained with (b) arterial gas saturation (SaO2) measurements pre contrast agent injection (c) post contrast injection (d) TRUST MRI vein oxygenation (SvO2) and (d) MRvF SO2. T-test were used for statistical analysis (**adapted from Christen 2018**[31])

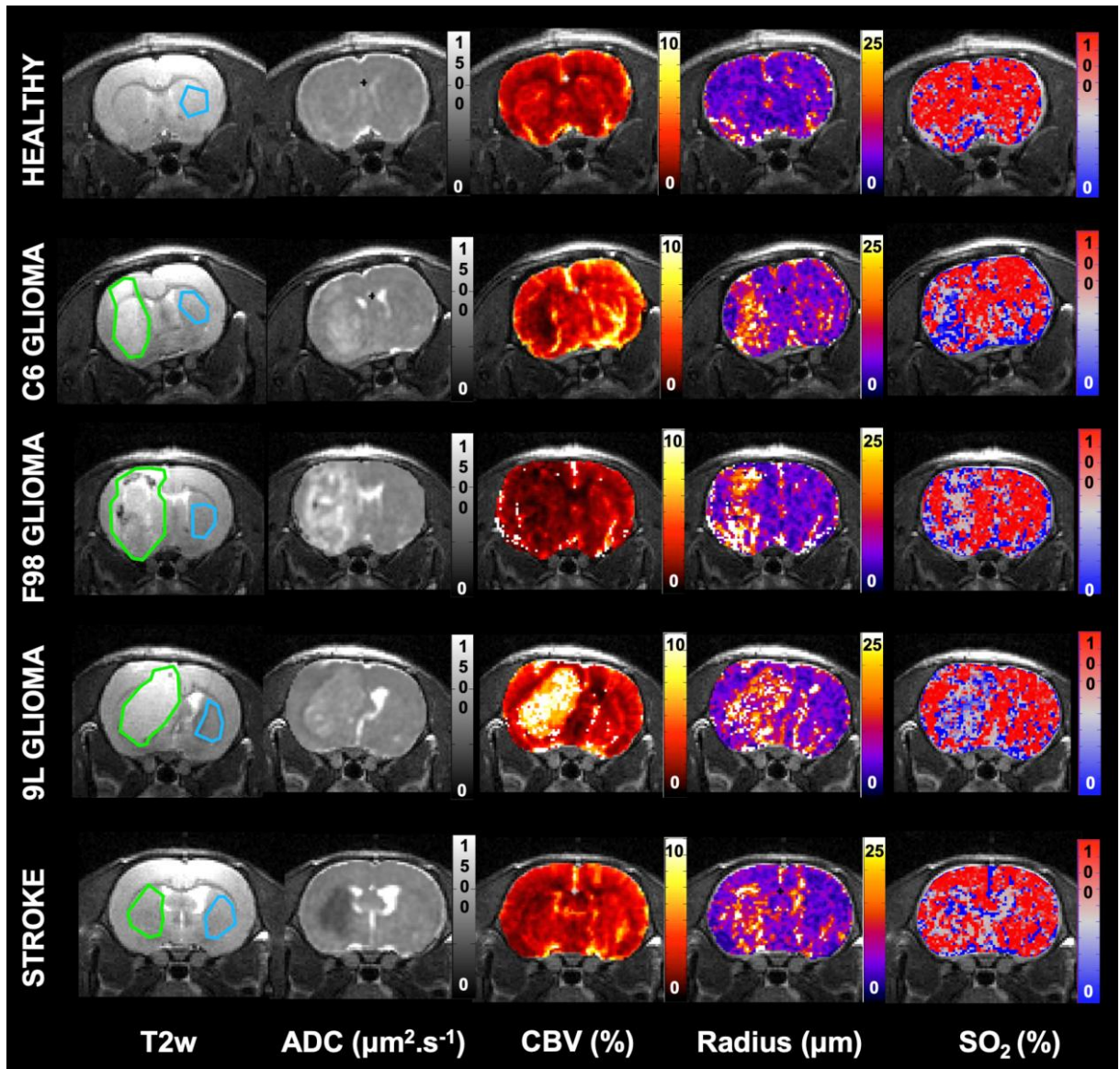

**Figure 5:** First MRvF results in animals including 3 glioma models and one model of stroke. (**adapted from Lemasson 2016**[33]). From left to right: anatomical T2w image for ROIs delineation, reference ADC map, and computed MRvF quantitative maps of CBV, R and SO2.

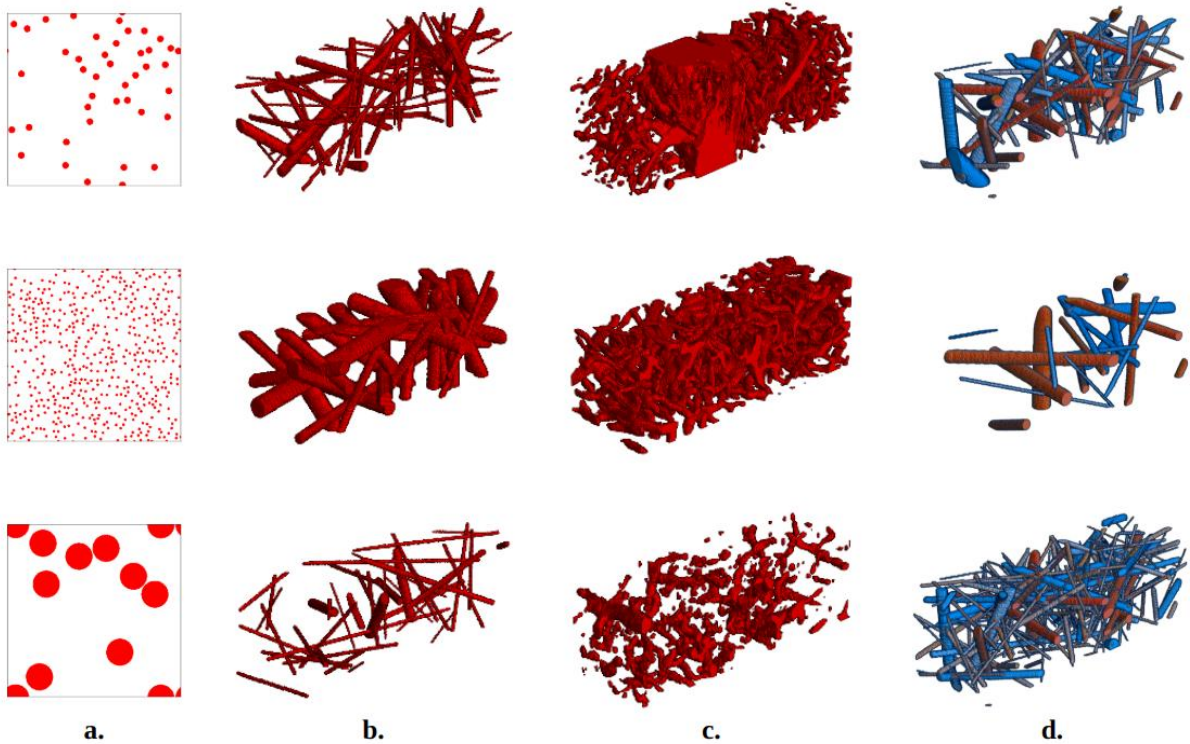

**Figure 6:** Illustrations of possible geometrical models to consider during MRvF dictionary generation. For each column, 3 examples with different blood volume and radii values are shown. (a) 2D model used in earlier studies. All vessels are orthogonal to the voxel plane. (b) 3D model using straight cylinders. (c) 3D Vascular network segmented from microscopies. (d) 3D model using straight cylinders, which each have their own SO2 value to mimic veins (in blue) and arteries (in red) in the voxel. **(adapted from Delphin 2024[42])**

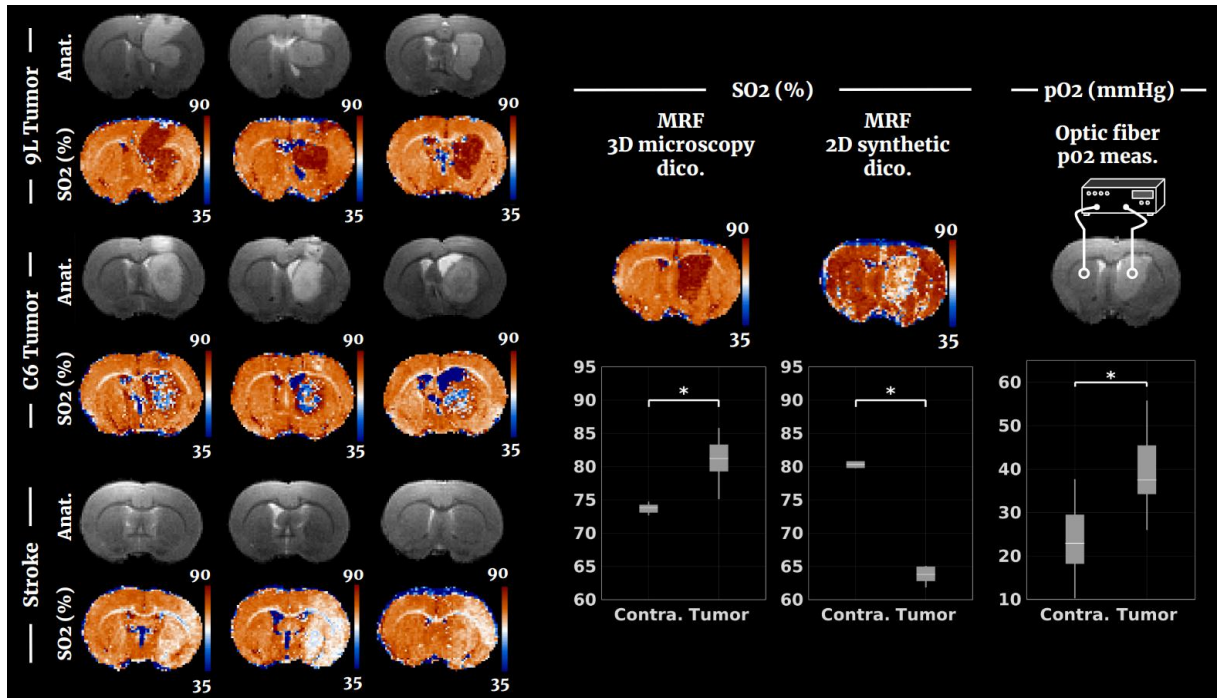

**Figure 7:** Illustrations of finding following the method proposed in **Delphin et al. 2024.** (left) SO2 maps obtained on rats with the dictionary based on 3D microscopy voxels with the corresponding anatomical images for reference. Three pathologies are shown here, the hyperoxic 9L tumor, the heterogeneous C6 tumor, and a stroke model. (right) Illustration of the difference in SO2 maps (%) obtained with a dictionary based on 3D microscopies and another one based on 2D synthetic voxels, along with the associated group statistics (lesion and contra ROIs across all animals, N=9). The group statistics is also presented for pO2 measurements in mmHg (N=6). A star indicates a p-value < 0.05 for a two-sample t-test. **(adapted from Delphin 2024[42])**

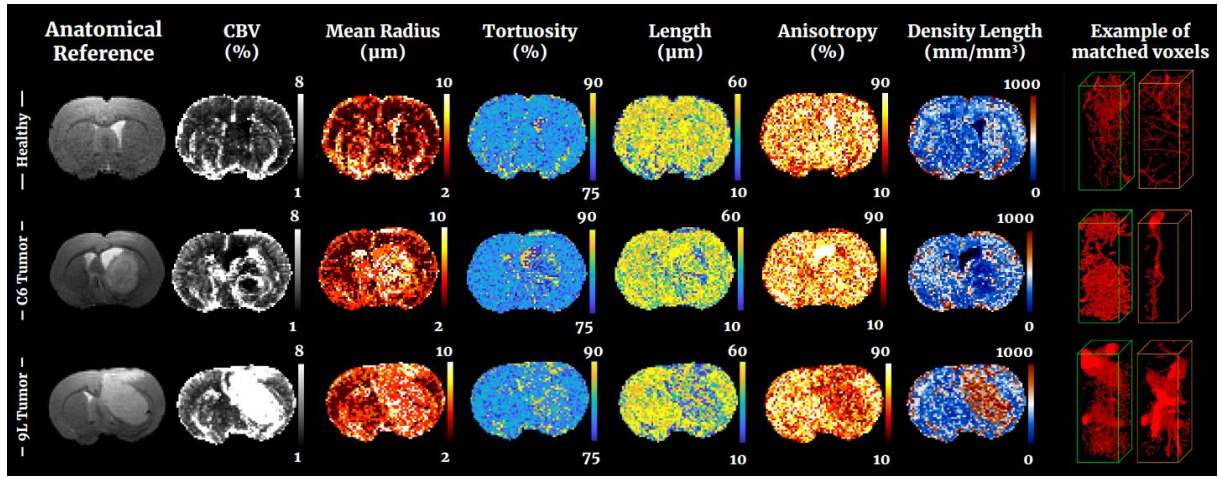

**Figure 8:** Illustration of the different maps of geometrical features that can be obtained with a dictionary based on 3D microscopy voxels with complex characterization, as detailed in Marçal et al. 2025. An example of a healthy rat is presented, along animals bearing C6 or 9L tumors. The rightmost column illustrates typical voxels that can be found in the cortex and striatum of the healthy animal, and in the tumors for the corresponding animals. **(adapted from Marcal 2025**[45]**)**

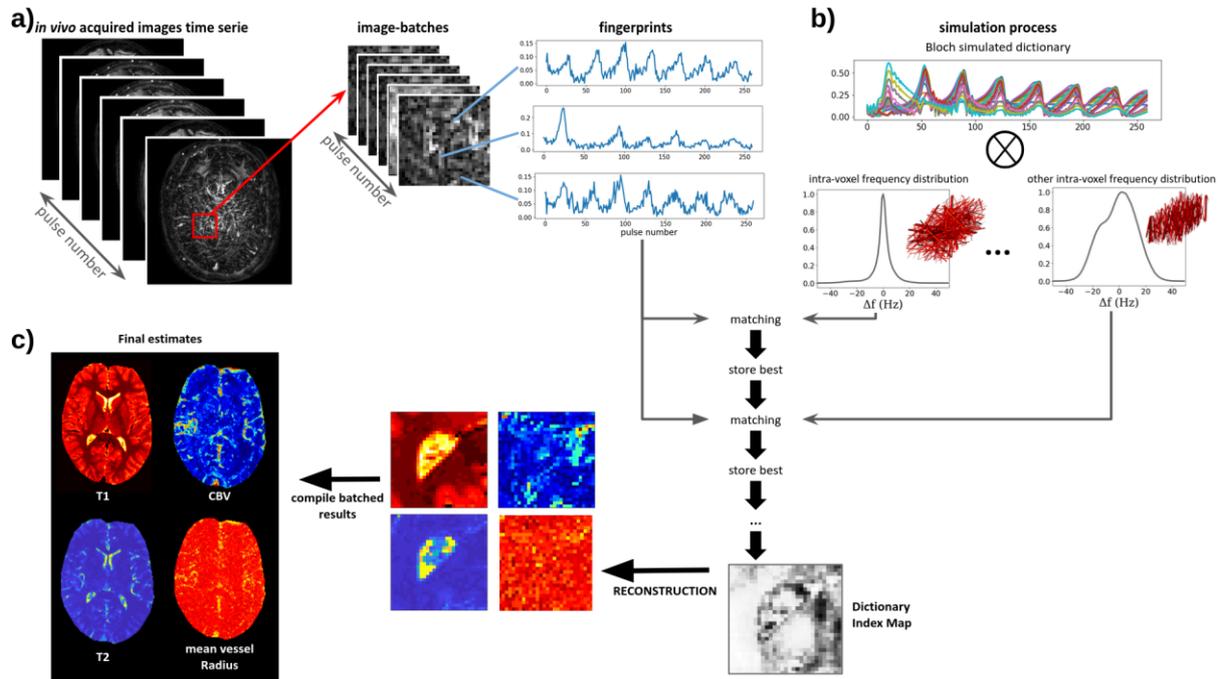

**Figure 9:** Schematic illustration of the on-the-fly matching/simulating process described in **Coudert et al 2025**[49] (a) in vivo acquisition images are batched and time-resolved fingerprints are extracted (b) in silico simulations are computed using few intra-voxel frequency distributions at the same time. The Bloch dictionary convolved with distributions is then used for matching with the fingerprints, resulting in an index map that is used for the reconstruction of (c) batched parameters maps that are compiled to retrieve full maps (with only 4 parameters over 6 as examples).

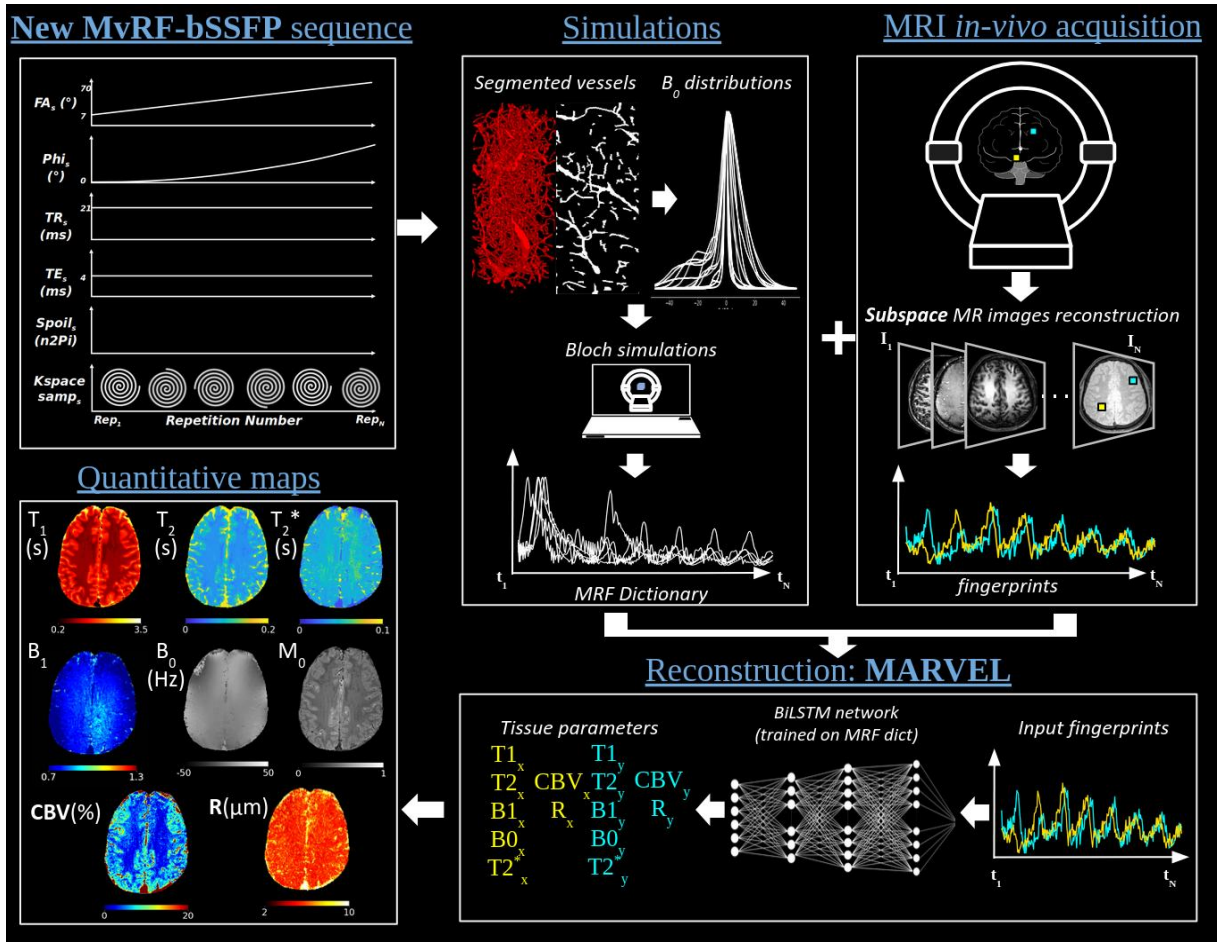

**Figure 10:** non-contrast MRvF framework. The Bloch simulation model uses realistic 3D vascular networks and a phase cycling bSSFP sequence (**Coudert 2025**[49]). A deep learning network (**Barrier 2024**[67]) is used to simultaneously reconstruct quantitative relaxometry and vascular maps. This replaces the matching process presented in Figure 1 and 2.

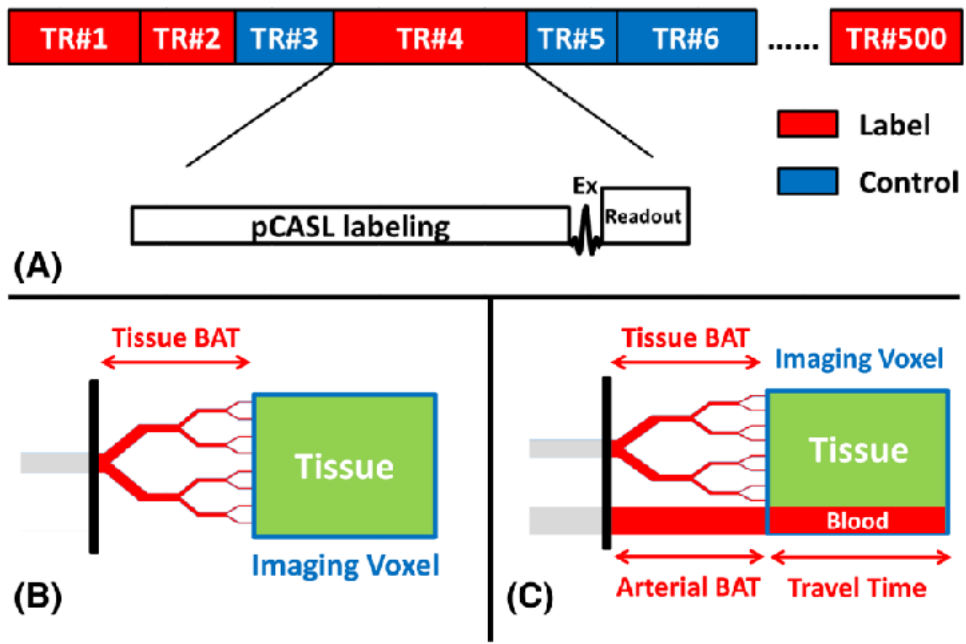

**Figure 11:** Pulse sequence and kinetic models of MRF-ASL. (a) Schematic diagram of the MRFASL pulse sequence. The sequence consists of varying duration, randomly ordered control and label pCASL modules, each followed immediately by an acquisition; (b) single-compartment model. The imaging voxel is assumed to contain tissue only; (c) two-compartment model. The voxel contains an additional compartment of passing-through arterial compartment. Reproduced with kind permission from **(Su 2020**[59]**)**

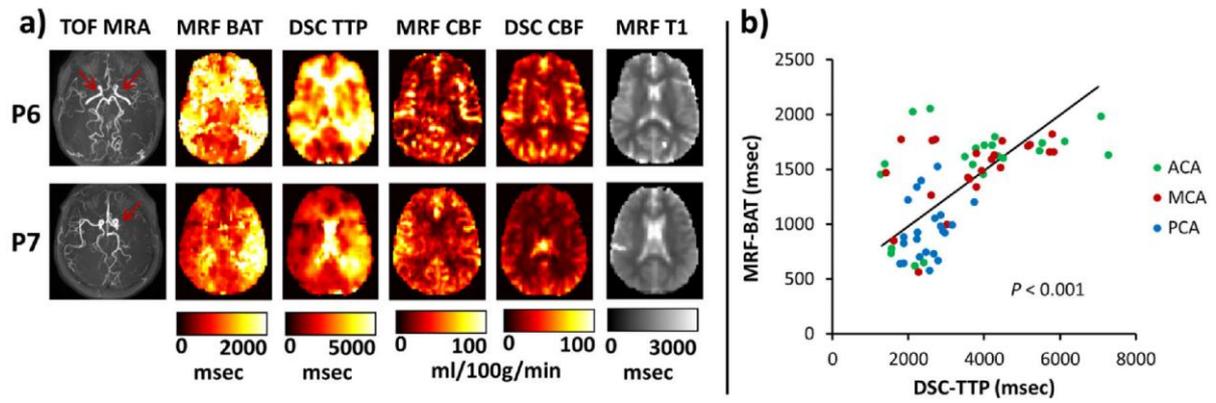

**Figure 12:** Comparison of MRF-ASL and DSC results in Moyamoya patients. (a) Parametric maps in two representative patients. Patient #6 had bilateral stenosis; Patient #7 had left MCA occlusion; (b) Scatter plot of MRF-ASL BAT and DSC TTP values in regions of interest. Each dot represents data from one perfusion territory (left and right ACA, MCA, and PCA) of one patient. Reproduced with kind permission from **(Su 2022**[62]**)**

| Authors | Number of parameters | Estimates | Acquisition | Number of signals | Model | Reconstruction | Application |
|---|---|---|---|---|---|---|---|
| Christen et al.[28] 2014 | 3 | CBV, R, SO2 | 3T GESFIDSE pre/post ratio | 52,920 | 2D synthetical | Standard matching | Healthy volunteers |
| Lemasson et al.[33] 2016 | 3 | CBV, R, SO2 | 4.7T GESFIDSE pre/post CA ratio | 38,976 | 2D synthetical | Standard matching | Tumor and stroke animal models |
| Pouliot et al.[41] 2017 | 5 | CBV, R, SO2, [SPION], ΔB0 | 7T GESFIDSE pre/post CA ratio | 404,000 | 3D mouse angiograms | Standard matching | Atherosclerotic animal model |
| Su et al.[58] 2017 | 4 | B1+, T1, CBF, BAT | 3T MRF-ASL | >1,000,000 | 1 compartment | Standard matching | Healthy volunteers Moyamoya patients |
| Su et al.[58] 2017 | 7 | B1+, T1, CBF, tissue BAT, arterial BAT, arterial CBV, tissue transit time | 3T MRF-ASL | 87,964,800 | 2 compartments | Standard matching | Healthy volunteers Moyamoya patients |
| Wright[60] 2018 | 6 | B1+, T1, CBF, tissue BAT, arterial BAT, arterial CBV | 3T MRF-ASL with variable PLD | 7200 / 4000 / 400 | 2 compartments | 3-step dictionary matching | Healthy volunteers |
| Wang et al.[11] 2019 | 4 | T1, T2, T2*, ΔB0 | 3T optimized bSSFP | 30,251,520 | T2* Lorentzian | Standard matching | Healthy volunteers |
| Zhang et al.[68] 2020 | 4 | B1+, T1, CBF, BAT | 3T optimized MRF-ASL | 6,149,000 | 1 compartment | Fully connected neural network | Healthy volunteers Moyamoya patients |
| Zhang et al.[68] 2020 | 7 | B1+, T1, CBF, tissue BAT, arterial BAT, arterial CBV, tissue transit time | 3T optimized MRF-ASL | 87,964,800 | 2 compartments | Fully connected neural network | Healthy volunteers Moyamoya patients |
| Fan et al.[69] 2021 | 4 | B1+, T1, CBF, BAT | 3T optimized MRF-ASL | ~1,000,000 | Training data acquired in vivo | Fully connected neural network | Healthy volunteers 1 Moyamoya patient |
| Su et al.[62] 2022 | 3 | BAT, CBF, T1 | 3T optimized MRF-ASL | >1,000,000 | 1 and 2 compartments | Standard matching | Moyamoya patients |
| Barrier et al.[67] 2024 | 6 | T1, T2, B1, ΔB0, CBV, R | 3T optimized balanced GRE sequence | >6e7 for sobol-distributed online training | 3D mouse microscopy | LSTM neural network | Healthy volunteers |
| Fan et al.[61] 2024 | 8 | T1, B1+, CBF, tissue ATT, ADC 3D, T2* | 3T variable-TE MRF-ASL with diffusion weighting | 6,149,000 / 373,248 / 94 | 1 and 2 compartments | 3-step dictionary matching | Healthy volunteers Stroke patients |
| Coudert et al.[49] 2025 | 6 | T1, T2, B1, ΔB0, CBV, R | 3T optimized balanced GRE sequence | 975,000,000 | 3D mouse microscopy | Online matching | Healthy volunteers |
| Oudoumanessah et al.[65] 2025 | 6 | T1, T2, B1, ΔB0, CBV, R | 3T optimized balanced GRE sequence | 975,000,000 | 3D mouse microscopy | Compressed HD-MED matching and GLLiM | Healthy volunteers |
| Marça et al. ISMRM[45] 2025 | 8 | CBV, R, SO2, T2, advanced voxels characteristics | 4.7T GESFIDSE pre/post CA concatenate | 135,000 | Augmented 3D mouse microscopy | Standard matching | Tumor animal model |

**Table 1**: Overview of major methodological MRvF studies. From right to left columns: authorship, Number of parameters: total MRvF parameters in the dictionary, Estimates: measured MRvF parameters, Acquisition: scanner field strength & sequence, Number of signals: number of dictionary

signal entries, Model: vascular geometry model used in simulations, Reconstruction: Parametric maps reconstruction method, and Application: study context.